\documentclass[sigconf]{acmart}

\AtBeginDocument{%
  }

\setcopyright{none}
\renewcommand\footnotetextcopyrightpermission[1]{}
\usepackage{booktabs}
\usepackage{xspace}
\usepackage[most]{tcolorbox}
\usepackage{enumitem}
\usepackage{xcolor}

\definecolor{background_u}{RGB}{245,250,245}
\definecolor{frame_green}{RGB}{60,120,80}

\newcommand{\AS}{\textsc{AgentSelect}\xspace}
\newcommand{\OC}{\textsc{ClawHub}\xspace}

\begin{document}

\title[Adapting Embedding Models for Agent Capability Retrieval]{Adapting Embedding Models for Agent Capability Retrieval}

\author{Tingwei Chen}
\email{Tingwei.Chen@student.uts.edu.au}
\affiliation{%
  \institution{University of Technology Sydney}
  \city{Sydney}
  \country{Australia}}

\author{Yunxiao Shi}
\authornote{Corresponding authors.}
\email{Yunxiao.Shi@student.uts.edu.au}
\affiliation{%
  \institution{University of Technology Sydney}
  \city{Sydney}
  \country{Australia}}

\author{Zhengdong Chu}
\email{zc9uy@virginia.edu}
\affiliation{%
  \institution{Squirrel AI Learning}
  \country{USA}}

\author{Qingsong Wen}
\email{qingsongedu@gmail.com}
\affiliation{%
  \institution{Squirrel AI Learning}
  \country{USA}}

\author{Min Xu}
\email{min.xu@uts.edu.au}
\affiliation{%
  \institution{University of Technology Sydney}
  \city{Sydney}
  \country{Australia}}

\renewcommand{\shortauthors}{Chen et al.}

\begin{abstract}
Open agent marketplaces list native agents, tool bundles, and reusable
skill packages in the same search interface, yet practitioners still have
little guidance on how to retrieve across this mixed catalog. We study whether off-the-shelf retrieval models, trained for general text
retrieval, can be adapted to match user queries to executable agent
capabilities, and whether the learned signal transfers beyond the benchmark
used for tuning. We fine-tune three open retrieval backbones, BGE-base,
KaLM-v1.5, and EasyRec, on \AS{}, which represents
marketplace-visible units as capability profiles derived from public metadata, and test transfer on two
catalogs not seen during training: MuleRun native agents and
a \OC{} benchmark of 50 skills with 1{,}000 queries.
Adaptation helps on both catalogs. Code and data will be released upon publication.
\end{abstract}


\maketitle

\section{Introduction}
LLM routing, tool retrieval, and skill discovery are usually studied
separately, but open marketplaces increasingly expose them through a single
search surface. At the interface, a query such as \emph{``transcribe this
audio with speaker labels''}, \emph{``automate the weekly invoice export from
Notion''}, or \emph{``scrape product reviews from a JS-rendered shop''} may
surface a deployable LLM agent, a function-calling tool bundle, or a
\texttt{SKILL.md}-based agent skill package. We use \emph{skill} to mean a reusable instruction package that a host LLM
can load on demand, sometimes together with scripts or helper tools. Some
skills ship concrete tools such as API clients, MCP servers, or browser
automation. Others are pure prompt templates with progressive disclosure. In
both cases, the marketplace item is better understood as a capability
than as a plain tool wrapper. These lines are usually studied on their own. LLM routing chooses a backbone
model~\cite{RouterEval,routellm,UniRoute,Arch-Router,OmniRouter,RouterBench};
tool retrieval picks a subset of a function-calling
inventory~\cite{ToolRet,RAG-MCP,MCP-Zero,MCPBench,efficient_tool_rep,tulip_agent};
skill discovery lives as a side-channel inside individual agent
frameworks~\cite{wang2026skillx,xu2026agent}. None of these splits is visible to the
end user, who simply wants whichever executable unit can do the job.

\begin{figure*}[t]
\centering
\includegraphics[width=\textwidth]{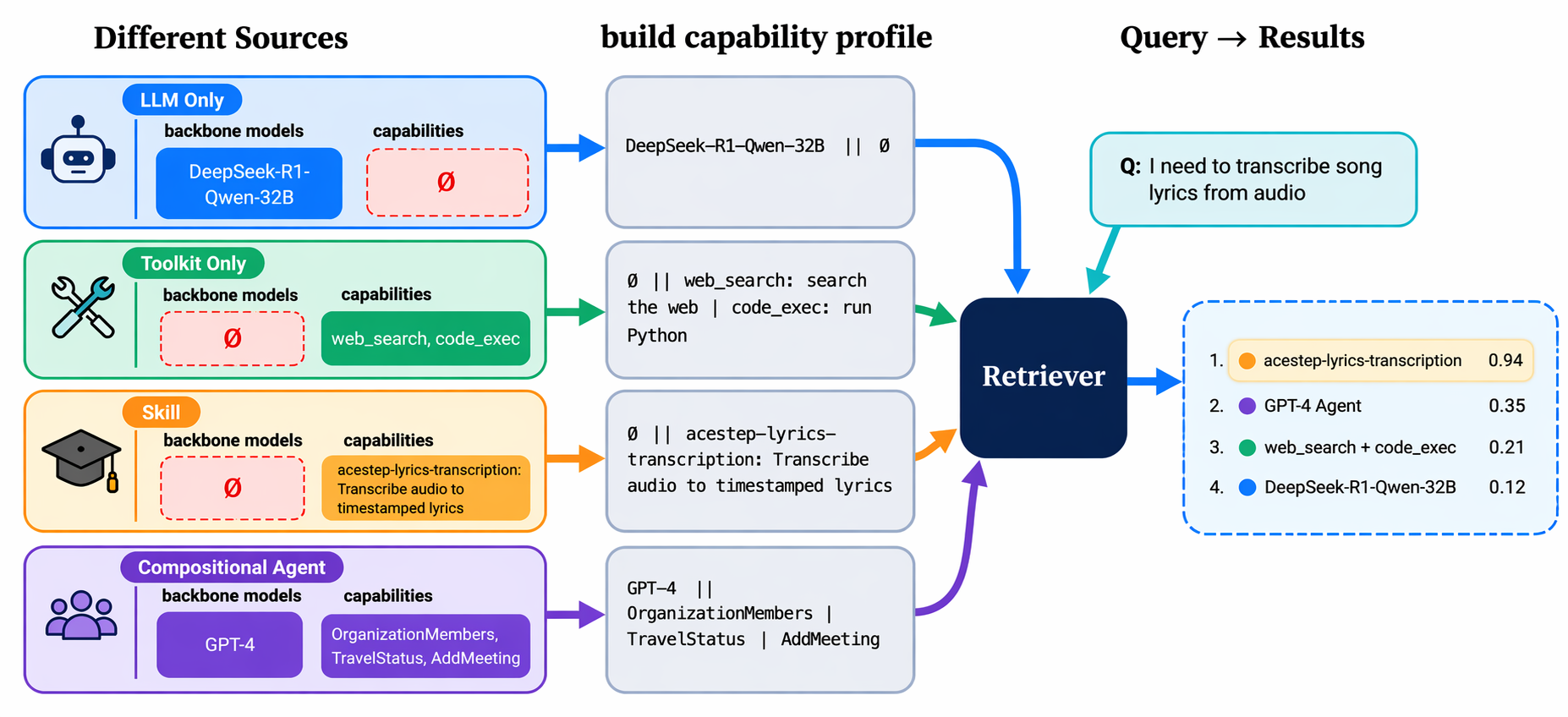}
\caption{AgentSelect benchmark profiles, MuleRun marketplace entries, and
ClawHub \texttt{SKILL.md}-based agent skill packages can all be serialized as capability
profiles and scored by the same retriever at the metadata retrieval stage. This paper then
tests whether training on the full \AS{} benchmark transfers to two
unseen public catalogs: MuleRun native agents and a 50-skill \OC{} benchmark.}
\label{fig:unification}
\end{figure*}

\noindent \textbf{Our scope.} The three lines differ in operational respects: routing
may also depend on cost or latency, tool selection may require subset
prediction, and skill execution depends on runtime and permission boundaries. Our claim is that they admit a shared
retrieval view under marketplace metadata. Once an
executable unit is written as a \emph{capability profile}, a short natural-language description of what it
can do, agents, tool bundles, and skills can
be placed in a common candidate catalog and scored by the same retrieval
interface against a narrative query. As Figure~\ref{fig:unification} shows, a skill's bundled tools and
instruction body form its capability profile, and a compatible host LLM
supplies the runtime. This keeps the differences among routing, tool use, and skill execution,
while giving a common representation for studying retrieval from public
catalog metadata alone.

If capability-level signals matter for retrieval, then supervision learned
from synthetic capability-profile interactions should transfer to unrelated
public catalogs of native marketplace agents and skills recast as
lightweight agents. We test this implication with three off-the-shelf
retrievers fine-tuned on \AS{}~\cite{shi2026agentselect}, and we evaluate on
two public catalogs the supervision never touched: MuleRun~\cite{mulerun}
native marketplace agents, and a \OC{} benchmark built from 50 selected
skills and 1{,}000 prompt-generated queries.
We make the following contributions:
\begin{itemize}
\item \textbf{A matched adaptation study for agent-capability retrieval.} We show
that three openly available retrieval models of different families and
parameter scales, BGE-base~\cite{bge_embedding,llm_embedder,cocktail} (BERT-base bi-encoder, 110M),
KaLM-v1.5~\cite{zhao2025kalmembeddingv2} (Qwen2-0.5B bi-encoder), and
EasyRec~\cite{ren2024easyrec} (RoBERTa-base 125M bi-encoder), can be
adapted under a shared LoRA protocol on \AS{} to improve
query-to-agent matching on the benchmark and, more importantly, on two
unseen public catalogs.
\item \textbf{A metadata-level representation.} We give an explicit
mapping in \S\ref{sec:problem} that places agents, tool bundles, and
skills into a shared capability-profile representation so they can be ranked
within one retrieval interface.
\item \textbf{Transfer evidence and a new evaluation benchmark.} The adapted
retrievers improve on MuleRun native marketplace agents and on \OC{}, a new
evaluation benchmark we construct from 50 public agent skills with 1{,}000
prompt-generated queries at five controlled difficulty levels. The
construction pipeline, including the prompt template and distractor-aware
generation strategy, is released for reuse.
\end{itemize}

\section{Related Work}
\label{sec:related}
\textbf{LLM routing.} Recent benchmarks and protocols evaluate cost--quality
trade-offs across language model
backbones~\cite{RouterEval,routellm,UniRoute,Arch-Router,OmniRouter,RouterBench}. They choose a backbone for the query but do not model a toolkit.

\noindent \textbf{Tool retrieval.} A second line retrieves function-calling tools from
large inventories given a
query~\cite{ToolRet,RAG-MCP,MCP-Zero,MCPBench,efficient_tool_rep,tulip_agent,toolbench}. They pick a subset of tools for the query, with the backbone fixed or
ignored.

\noindent \textbf{Agents and skill libraries.} Agent frameworks
(e.g., AutoAgent~\cite{AutoAgent})
expose components for assembling agents on the fly.
A growing body of work treats skills as first-class reusable
units~\cite{xu2026agent}: Voyager~\cite{voyager} grows a library of
executable skills retrieved by description,
SkillX~\cite{wang2026skillx} automatically constructs hierarchical skill
knowledge bases, and Anthropic's Claude Agent
Skills~\cite{anthropic_agent_skills} provide one recent \texttt{SKILL.md}-based implementation, with
metadata, an instruction body, tool bundles, and progressive disclosure. Public registries such as
ClawHub~\cite{clawhub} make thousands of skills discoverable.

\noindent \textbf{Agent recommendation.} \AS{}~\cite{shi2026agentselect} (the
unified-supervision benchmark on which we build) treats narrative
query-to-agent recommendation over capability profiles as one task;
iAgent~\cite{iagent} and Agent4Rec~\cite{Agent4Rec} target user-modelling
proxies, which is a different problem. SkillOrchestra~\cite{wang2026skillorchestra}
learns to route among agents by transferring skills, and LTRR~\cite{kim2026ltrr}
learns to rank retrievers per query; both take the candidate agents or
retrievers as given rather than adapting a single retriever to a shared
capability-profile representation. To our knowledge, no prior work
places agent retrieval, tool retrieval, and skill retrieval in a single
retrieval framework with a single capability-profile representation.

\noindent \textbf{Why these lines are related.} Each of the three lines above implicitly
operates on a strict subset of the capability profile: routing fixes the
toolkit to empty, tool retrieval fixes the backbone, and skill discovery
fixes both at the host-LLM defaults of a particular framework. Reading them this way, our
proposal is not that they become identical tasks, but that they admit a
shared retrieval interface over marketplace-visible capability metadata. This
is the hypothesis we evaluate in the transfer experiments below.

\section{Preliminaries}
\label{sec:problem}
Prior work~\cite{shi2026agentselect} formalizes a marketplace entry as a
pair of a backbone model and a toolkit, and phrases query-to-agent
recommendation as ranking such pairs against a narrative query. In practice,
marketplace metadata is heterogeneous: some entries expose an explicit
backbone, others do not, and many share the same default host LLM. We
therefore adapt the setting to a more direct form and represent each entry
by its \emph{capability profile}: a short natural-language view drawn
directly from the public metadata, including backbone and tool information
when available. Native agents, tool bundles, and
\texttt{SKILL.md}-based skills then live in a single catalog that one
retriever can rank against a narrative query.

\noindent \textbf{Skills as lightweight agents.} A \texttt{SKILL.md}-based
agent skill is a self-contained markdown package with
(i)~name and description metadata, (ii)~an instruction body, and (iii)~a
bundle of tools backed by APIs, CLIs, MCP servers, browser automation, or
database access. Writing the skill as a capability profile keeps all three
components as natural-language text, placing skills in the same catalog
as native agents and tool bundles.

\noindent \textbf{Scope of the abstraction.} The capability profile is the
common core directly visible in marketplace metadata and rich enough to
distinguish many executable units. Empirically a name-plus-description view
already captures the matching signal: in our \OC{} ablation
(Table~\ref{tab:ablation-repr}), it matches a full-field representation
that appends every \texttt{SKILL.md} field while using $2.6\times$ fewer
tokens. Other deployment factors such as cost, latency, safety, host
runtime, memory, and permission boundaries are not encoded by the profile;
the abstraction targets the case where the retrieval system only sees
public catalog metadata.

\section{Supervision and Models}
\label{sec:method}
Our procedure takes three off-the-shelf retrievers and fine-tunes each on the
full \AS{} benchmark: \textbf{BGE-base}~\cite{bge_embedding,llm_embedder,cocktail}, the
\textbf{KaLM-v1.5} embedding model~\cite{zhao2025kalmembeddingv2}, and the
recommendation-tuned \textbf{EasyRec} encoder~\cite{ren2024easyrec}. These three cover a BERT-base bi-encoder, an LLM-based embedder, and a
recommendation-tuned encoder at different scales, so a consistent adaptation
effect is less likely to come from one architecture. We restrict the study to dense bi-encoders and leave sparse
(SPLADE) and late-interaction (ColBERT) retrievers to future work. We mark adapted variants with a
star ($^{*}$). Supervision is positive-only query--profile interaction data;
negatives are sampled in-batch. All models use the same textual serialization
of the available profile fields: the backbone identifier when exposed, plus
tool names and tool descriptions, are converted into a single text view of
the profile. No
architectural change is made to any of the three models. In the analysis we
highlight Part~II and Part~III because they are the benchmark subsets most
relevant to toolkit-centric and compositional capability matching, while the
full benchmark results are reported in \cite{shi2026agentselect}. The same three backbones are then evaluated on the unseen MuleRun and \OC{} catalogs.

\noindent \textbf{Training setup.} All three backbones are fine-tuned with LoRA
adapters under the same optimizer, loss, and data pipeline. Common settings: AdamW, 10 epochs,
in-batch contrastive loss with temperature $0.05$ plus one random negative
per positive.
Table~\ref{tab:lora-recipe} reports the final per-backbone configuration.

\noindent \textbf{Model selection.} For each transfer target (MuleRun and
\OC{}), we hold out 10\% of the queries at random as a validation split and
use the remaining 90\% as the test split. After every epoch, each adapter is
evaluated on both the validation and test splits of both targets. We select
the best checkpoint independently per target by validation nDCG@10 and report
the corresponding test-split metrics. This yields a separate best epoch per
model--target pair (e.g.\ BGE-base selects epoch~3 for MuleRun and epoch~6
for \OC{}).

\begin{table}[t]
\caption{LoRA configuration per backbone.}
\label{tab:lora-recipe}
\centering
\small
\setlength{\tabcolsep}{4pt}
\begin{tabular}{lccccc}
\toprule
Backbone & Size & LR & Rank / $\alpha$ & Target & \% Train. \\
\midrule
EasyRec      & 125M & 5e-6 & 8 / 16 & q,k,v (last 4 layers) & 0.12 \\
BGE-base    & 110M & 1e-5 & 8 / 16 & q,k,v (last 4 layers) & 0.13 \\
KaLM-v1.5    & 0.5B & 1e-5 & 8 / 16 & q\_proj, v\_proj (all) & 0.11 \\
\bottomrule
\end{tabular}
\end{table}

\section{Datasets}
\label{sec:datasets}
\AS{}~Part~II provides toolkit-centric supervision: each query is paired with
a reference toolkit~\cite{toolbench}. \AS{}~Part~III provides synthesized
compositional capability profiles whose pseudo-positives are validated by
counterfactual capability edits~\cite{shi2026agentselect}.

MuleRun~\cite{mulerun} is a public agent marketplace. Following~\cite{shi2026agentselect}, we report the final filtered evaluation
subset: 59 native agents and 1{,}080 narrative queries. Each agent is mapped
into the toolkit-only schema by inferring 5--10 tool primitives from its
marketplace description; no backbone identifier is recorded.

\OC{}~\cite{clawhub} is a public marketplace for agent skills. We build a transfer evaluation benchmark as follows.

\noindent \textbf{Skill selection.} From the marketplace we keep 50 skills
whose \texttt{SKILL.md} pages describe a concrete, testable capability,
discarding writing-persona and style-only entries whose target behavior is
too diffuse to serve as retrieval targets. Each retained skill is parsed
into a capability profile that summarizes its name, intended use, and tool
bundle.

\noindent \textbf{Distractor pool.} For each target we treat the
profiles of the remaining skills in the catalog as \emph{distractors}
and pass them alongside the target profile. Because the 50 skills span a
narrow band of marketplace capabilities, these distractors already provide
vocabulary and adjacent-capability overlap with the target, so we use the
full pool rather than a cluster-sampled subset.

\noindent \textbf{Narrative query generation.} For each target, a single
prompt conditioned on the target profile and the distractor profiles
elicits 20 narrative queries from GPT-5.4. The prompt
(Figure~\ref{fig:query-prompt}) asks for an approximately uniform spread
across the five prompt levels L1--L5 of
Table~\ref{tab:difficulty-ladder} (illustrated for a single intent in
Table~\ref{tab:difficulty-examples}) together with an intent-type mix
(explicit, semi-explicit, implicit, scenario-driven, constraint-first,
noisy, ambiguous), and requires lexical, syntactic, and structural
diversity. The model returns a JSON array that also records each query's
level label, an intent-type tag, a rationale, and a short note on whether
distractor semantics shaped the phrasing.

\noindent \textbf{Assembly.} The 50 targets are processed independently in
parallel and their outputs concatenated; after removing near-duplicates, we
retain the final 50\,$\times$\,20\,=\,1{,}000 queries. We retain the level label and
distractor provenance for later analysis.

\begin{table}[t]
\caption{Five prompt levels (L1--L5) used for each \OC{} target.}
\label{tab:difficulty-ladder}
\centering
\small
\setlength{\tabcolsep}{4pt}
\begin{tabular}{p{0.08\linewidth}p{0.82\linewidth}}
\toprule
Level & Description \\
\midrule
L1 & Straightforward, explicit, clean requests.\\
L2 & Requests with light context or small constraints. \\
L3 & Requests with implicit goals or scenario framing. \\
L4 & Noisy, underspecified, or partially ambiguous requests. \\
L5 & Mixed signals, colloquial phrasing, irrelevant details, or soft
     confusion with distractor capabilities.\\
\bottomrule
\end{tabular}
\end{table}

\begin{table}[t]
\caption{Five prompt levels L1--L5 for a single intent (delete promotional/junk
email) on the \texttt{cedarscy-gmail-cleaner} skill.}
\label{tab:difficulty-examples}
\centering
\small
\setlength{\tabcolsep}{4pt}
\begin{tabular}{p{0.05\linewidth}p{0.86\linewidth}}
\toprule
Level & Query \\
\midrule
L1 & Trash promotional emails older than 30 days in my personal Gmail. \\
L2 & Without touching unread mail from the last 7 days, clear out the old promo stuff in Gmail. \\
L3 & Could use an inbox cleanup pass that surfaces the worst repeat senders first, then lets me wipe the obvious junk. \\
L4 & Need one of those Gmail cleanup things that can chew through the promo swamp and dump the ancient sale emails without touching the useful stuff. \\
L5 & I'm comparing agents and don't need web research or prompt polishing or anything fancy, just the one that can inspect my inbox, show the top senders, and bulk-trash the junk from them. \\
\bottomrule
\end{tabular}
\end{table}

\begin{figure*}[!p]
\begin{tcolorbox}[colback=background_u, colframe=frame_green, boxrule=0.5pt,
  title={\textbf{Query-generation prompt for one \OC{} target skill}},
  fontupper=\scriptsize,
  left=6pt, right=6pt, top=4pt, bottom=4pt]

\noindent You are a senior benchmark data curator for narrative-query-to-agent
recommendation.

\smallskip
\noindent Your task is to generate high-quality user narrative queries that
would plausibly be used to request help from a TARGET AGENT, given its
capability profile. You will be provided with: (1)~one TARGET AGENT profile,
and (2)~DISTRACTOR AGENT profiles drawn from the rest of the catalog. The distractor agents are NOT the
intended target, but they may be partially related and should be used as
semantic noise when constructing some ambiguous or weakly specified queries.

\smallskip
\noindent Your goal is to generate realistic, diverse, high-quality narrative
queries that:
\begin{itemize}[leftmargin=1.2em,itemsep=0pt,topsep=2pt]
  \item are primarily aligned with the TARGET AGENT's actual capabilities,
  \item vary from explicit intent to implicit intent,
  \item exhibit a gradual difficulty spectrum,
  \item include some realistic noise, ambiguity, redundancy, or colloquial phrasing,
  \item remain plausible as real user requests in an agent marketplace or application setting.
\end{itemize}
The generated queries should NOT read like annotation labels, feature
summaries, or capability checklists. They should read like natural user
requests.

\medskip
\noindent\textbf{Task definition.}\quad Generate \texttt{\{N\}} user
narrative queries for the TARGET AGENT. These queries should simulate
realistic user requests and should be suitable for benchmark construction
in an agent recommendation setting.

\medskip
\noindent\textbf{Input agents.}
\begin{description}[leftmargin=1.2em,itemsep=0pt,topsep=2pt,style=nextline]
  \item[\texttt{[TARGET AGENT]}] \texttt{\{target\_agent\_profile\}}
  \item[\texttt{[DISTRACTOR AGENTS]}] \texttt{\{distractor\_agent\_profiles\}}
\end{description}

\medskip
\noindent\textbf{Core requirements.}
\begin{enumerate}[leftmargin=1.4em,itemsep=1pt,topsep=2pt,label=\arabic*.]
  \item \emph{Target alignment.} Each query should be meaningfully solvable
    by the TARGET AGENT. The query should reflect the target agent's
    capability boundaries faithfully. Do not introduce requirements that
    clearly exceed the target agent's described abilities.
  \item \emph{Distractor awareness.} The distractor agent profiles are
    provided as semantic noise. You may use them to inspire vague,
    partially underspecified, weakly disambiguated, mixed-intent, or
    marketplace-style user expressions that could superficially resemble
    other agents. However, the final query should still be more
    appropriate for the TARGET AGENT than for the distractors.
  \item \emph{Diversity of intent expression.} The query set must cover a
    spectrum from highly explicit, semi-explicit, implicit,
    scenario-driven, goal-first, constraint-first, noisy conversational,
    to marketplace search-like requests. Do NOT make all queries begin
    with patterns like ``I need an agent that can\ldots'', ``Help
    me\ldots'', or ``Can you\ldots''. Use diverse sentence structures.
  \item \emph{Difficulty ladder.} The set should show a gradual increase in
    difficulty:
    \begin{itemize}[leftmargin=1.2em,itemsep=0pt,topsep=1pt]
      \item Level~1: straightforward, explicit, clean requests.
      \item Level~2: requests with light context or small constraints.
      \item Level~3: requests with implicit goals or scenario framing.
      \item Level~4: noisy, underspecified, or partially ambiguous requests.
      \item Level~5: harder cases with mixed signals, colloquial language,
        extra irrelevant details, or soft confusion with distractor
        capabilities.
    \end{itemize}
    Difficulty should increase across the set rather than being random.
  \item \emph{Realistic noise.} Some queries should contain realistic
    user-side noise such as casual or colloquial language, incomplete
    specification, redundant context, self-correction, mild vagueness,
    platform-specific goals (e.g., social post, profile refresh, event
    poster, quick share), imprecise terminology, or slightly messy
    wording. But do NOT make them nonsensical or impossible to
    understand.
  \item \emph{Capability-faithfulness.} Do not hallucinate capabilities.
    If the target agent only performs a narrow transformation, the query
    should stay within that scope. Do not implicitly require additional
    design, editing, planning, or reasoning abilities unless they are
    clearly supported by the target profile.
  \item \emph{Real user style.} Queries should sound like what a real
    user would type in a marketplace, app, tool interface, or search
    box. Avoid overly technical, taxonomic, or benchmark-like wording.
    Avoid repetitive jargon such as ``identity-preserving pose
    variation'', ``micro-pose drift'', or ``semantic transformation''
    unless such language would genuinely be natural for a user.
  \item \emph{Lexical and structural diversity.} Ensure broad diversity
    in wording, syntax, narrative framing, intent expression,
    explicitness level, and scene or application context. Avoid template
    repetition and Cartesian-product-like generation.
  \item \emph{Plausible ambiguity.} A subset of queries may be
    intentionally a bit ambiguous or weakly specified, especially when
    distractor profiles make that ambiguity realistic. However:
    ambiguity should be natural, the TARGET AGENT should still be the
    best or one of the best matches, and the query should not become
    equally or more suitable for a distractor.
  \item \emph{Benchmark value.} The final set should be useful for
    evaluating semantic understanding, intent matching, robustness to
    paraphrase, robustness to user noise, implicit intent recognition,
    and boundary-sensitive recommendation.
\end{enumerate}

\medskip
\noindent\textbf{Negative constraints.}\quad Do NOT: produce
near-duplicate paraphrases; produce rigid template-based rewrites; write
capability summaries instead of user requests; overuse ``I need an agent
that can\ldots''; make every query equally explicit; make every query
perfectly clean and polished; include impossible requirements
unsupported by the target agent; overfit to exact profile wording; or
copy phrases from the agent description verbatim unless naturally
needed. Do NOT make the generated data feel synthetic through obvious
templating.

\medskip
\noindent\textbf{Generation strategy.}\quad Before writing the final
queries, internally reason about the core capability kernel of the
TARGET AGENT, its capability boundaries, which distractors are closest
and could create realistic ambiguity, which user intents are direct vs.\
indirect, and which contexts are plausible but still
capability-faithful. Then generate the queries accordingly.

\medskip
\noindent\textbf{Output format.}\quad Return a JSON array. Each item
must contain:
\begin{itemize}[leftmargin=1.2em,itemsep=0pt,topsep=2pt]
  \item \texttt{id}: integer index starting from 1.
  \item \texttt{query}: the narrative query.
  \item \texttt{difficulty}: one of \texttt{["L1","L2","L3","L4","L5"]}.
  \item \texttt{intent\_type}: one of \texttt{["explicit","semi\_explicit","implicit","scenario\_driven","constraint\_first","noisy","ambiguous"]}.
  \item \texttt{rationale}: a brief explanation of why this query fits
    the TARGET AGENT and what makes it distinct.
  \item \texttt{noise\_from\_distractors}: brief note on whether and how
    distractor-agent semantics weakly influenced the phrasing.
\end{itemize}
Ensure the final output is valid JSON only.

\medskip
\noindent\textbf{Distribution constraints.}\quad Generate exactly
\texttt{\{N\}} queries with approximate distribution of 4 queries per
level L1--L5. Intent-type coverage: at least 2 explicit or
semi\_explicit, at least 2 implicit or scenario\_driven, at least 2
noisy, at least 2 ambiguous or weakly specified, and at least 1
constraint\_first. Diversity: no two queries should share the same
opening structure; no more than 2 queries may explicitly mention the
agent output format if the user would not naturally say it; at least 3
queries should frame the need through a real-life use case rather than
direct capability description; at least 3 queries should contain mild
noise or colloquial phrasing.
\end{tcolorbox}
\caption{Prompt template used to generate narrative queries per \OC{}
skill. At generation time, \texttt{\{N\}} is set to 20,
\texttt{\{target\_agent\_profile\}} is filled with the target's parsed
\texttt{SKILL.md} summary, and \texttt{\{distractor\_agent\_profiles\}}
is filled with the concatenated profiles of the other skills in the catalog.}
\label{fig:query-prompt}
\end{figure*}

\noindent \textbf{Level controls.} The labels L1--L5 come from the generation
prompt and are emitted in the JSON field named \texttt{difficulty}, but we do
not treat them as human-validated difficulty labels. Query length grows with
the level index, from 24.6 tokens on average for L1+L2 to 37.1 tokens for
L4+L5, because the later levels add scenario framing and
distractor-influenced context. At the same time, retrieval accuracy is not
monotonic in the label for all models: in
Table~\ref{tab:clawhub-difficulty}, some curves dip at L3 and partially
recover at L4 or L5. We therefore use L1--L5 as controlled prompt
conditions for query style rather than as a single scalar measure of
retrieval difficulty.

\section{Experiments}
\label{sec:experiments}
We first evaluate whether \AS{} supervision improves generic retrievers on
the benchmark subsets where toolkit and composition matter, and then whether
the gain transfers to the unseen MuleRun and \OC{} catalogs.

\noindent \textbf{Metrics.} Following~\cite{shi2026agentselect}, we report
P/R/F1/nDCG/MRR at cutoffs 1, 5, and 10 for MuleRun and \OC{}, while
benchmark results are discussed through the standard top-10 ranking metrics.
Because each query has a single gold agent, all five metrics at cutoff 1
reduce to Hit@1. In the discussion we therefore emphasize \textbf{P@1} and
\textbf{MRR@10}, and keep the full tables for comparability with prior work.

\subsection{In-domain adaptation on benchmark subsets}
The \AS{} benchmark paper~\cite{shi2026agentselect} already reports the full in-domain
leaderboards, so we restate only the subsets most relevant to the transfer question. When trained on the full
benchmark, all three retriever families improve on Part~II and Part~III; for
example, EasyRec improves from nDCG/MRR of 0.3494/0.2862 to 0.8552/0.8193 on
Part~II and from 0.3115/0.3105 to 0.6320/0.6501 on Part~III, with analogous
gains for BGE-base and KaLM-v1.5 reported in~\cite{shi2026agentselect}. We therefore treat the in-domain benchmark as
established evidence and use the remainder of this paper to focus on
transfer to unseen public catalogs.

\subsection{Transfer to unseen catalogs}
We next study whether the benchmark-trained signal transfers to external
catalogs the supervision never touched. All three adapted backbones are
re-evaluated on MuleRun and \OC{}. In every transfer table, rows without a
star are the off-the-shelf backbones evaluated zero-shot on the target
catalog, so each (base, starred) pair isolates the effect of \AS{} adaptation
over the model's own zero-shot baseline.

\noindent \textbf{MuleRun.} On 59 public marketplace agents with 1{,}080 queries
(972 in the test split),
all three bi-encoders gain from adaptation (Table~\ref{tab:rq23-transfer}). P@1 rises from 0.6296 to 0.7305
for EasyRec, from 0.5010 to 0.6409 for KaLM, and from 0.7212 to 0.7778 for
BGE-base. The same pattern appears in nDCG@10 and MRR@10.

\noindent \textbf{\OC{}.} The \OC{} benchmark covers 50 skills with prompt-generated queries.
Adaptation yields gains on all three backbones on the 900-query test split
(Table~\ref{tab:rq23-clawhub}). P@1 rises from 0.4989 to
0.5956 for EasyRec, from 0.4528 to 0.5611 for KaLM, and from 0.6667 to
0.7700 for BGE-base. The same pattern appears in nDCG@10 and MRR@10.

\begin{table}[t]
\caption{Transfer on MuleRun (test split, 972 queries). $^{*}$ denotes
tuning on the full \AS{} benchmark.}
\label{tab:rq23-transfer}
\centering
\small
\begin{tabular}{llccccc}
\toprule
Cutoff & Method & P & R & F1 & nDCG & MRR \\
\midrule
@1  & EasyRec       & 0.6296 & 0.6296 & 0.6296 & 0.6296 & 0.6296 \\
    & EasyRec$^{*}$ & \textbf{0.7305} & \textbf{0.7305} & \textbf{0.7305} & \textbf{0.7305} & \textbf{0.7305} \\
    & BGE-base     & 0.7212 & 0.7212 & 0.7212 & 0.7212 & 0.7212 \\
    & BGE-base$^{*}$ & \textbf{0.7778} & \textbf{0.7778} & \textbf{0.7778} & \textbf{0.7778} & \textbf{0.7778} \\
    & KaLM          & 0.5010 & 0.5010 & 0.5010 & 0.5010 & 0.5010 \\
    & KaLM$^{*}$    & \textbf{0.6409} & \textbf{0.6409} & \textbf{0.6409} & \textbf{0.6409} & \textbf{0.6409} \\
@5  & EasyRec       & 0.1603 & 0.8014 & 0.2671 & 0.7257 & 0.7001 \\
    & EasyRec$^{*}$ & \textbf{0.1749} & \textbf{0.8745} & \textbf{0.2915} & \textbf{0.8127} & \textbf{0.7917} \\
    & BGE-base     & 0.1716 & 0.8580 & 0.2860 & 0.7976 & 0.7772 \\
    & BGE-base$^{*}$ & \textbf{0.1741} & \textbf{0.8704} & \textbf{0.2901} & \textbf{0.8316} & \textbf{0.8183} \\
    & KaLM          & 0.1444 & 0.7222 & 0.2407 & 0.6269 & 0.5945 \\
    & KaLM$^{*}$    & \textbf{0.1568} & \textbf{0.7840} & \textbf{0.2613} & \textbf{0.7197} & \textbf{0.6981} \\
@10 & EasyRec       & 0.0862 & 0.8621 & 0.1568 & 0.7454 & 0.7083 \\
    & EasyRec$^{*}$ & \textbf{0.0897} & \textbf{0.8971} & \textbf{0.1631} & \textbf{0.8199} & \textbf{0.7945} \\
    & BGE-base     & 0.0882 & 0.8817 & 0.1603 & 0.8055 & 0.7806 \\
    & BGE-base$^{*}$ & \textbf{0.0888} & \textbf{0.8879} & \textbf{0.1614} & \textbf{0.8374} & \textbf{0.8208} \\
    & KaLM          & 0.0783 & 0.7829 & 0.1423 & 0.6466 & 0.6027 \\
    & KaLM$^{*}$    & \textbf{0.0832} & \textbf{0.8323} & \textbf{0.1513} & \textbf{0.7351} & \textbf{0.7043} \\
\bottomrule
\end{tabular}
\end{table}

\begin{table}[t]
\caption{Transfer on \OC{} (test split, 50 skills, 900 queries).}
\label{tab:rq23-clawhub}
\centering
\small
\begin{tabular}{llccccc}
\toprule
Cutoff & Method & P & R & F1 & nDCG & MRR \\
\midrule
@1  & EasyRec       & 0.4989 & 0.4989 & 0.4989 & 0.4989 & 0.4989 \\
    & EasyRec$^{*}$ & \textbf{0.5956} & \textbf{0.5956} & \textbf{0.5956} & \textbf{0.5956} & \textbf{0.5956} \\
    & BGE-base     & 0.6667 & 0.6667 & 0.6667 & 0.6667 & 0.6667 \\
    & BGE-base$^{*}$ & \textbf{0.7700} & \textbf{0.7700} & \textbf{0.7700} & \textbf{0.7700} & \textbf{0.7700} \\
    & KaLM          & 0.4528 & 0.4528 & 0.4528 & 0.4528 & 0.4528 \\
    & KaLM$^{*}$    & \textbf{0.5611} & \textbf{0.5611} & \textbf{0.5611} & \textbf{0.5611} & \textbf{0.5611} \\
@5  & EasyRec       & 0.1567 & 0.7833 & 0.2611 & 0.6528 & 0.6091 \\
    & EasyRec$^{*}$ & \textbf{0.1742} & \textbf{0.8711} & \textbf{0.2904} & \textbf{0.7480} & \textbf{0.7066} \\
    & BGE-base     & 0.1798 & 0.8989 & 0.2996 & 0.7939 & 0.7586 \\
    & BGE-base$^{*}$ & \textbf{0.1916} & \textbf{0.9578} & \textbf{0.3193} & \textbf{0.8767} & \textbf{0.8491} \\
    & KaLM          & 0.1462 & 0.7311 & 0.2437 & 0.5729 & 0.5198 \\
    & KaLM$^{*}$    & \textbf{0.1600} & \textbf{0.8000} & \textbf{0.2667} & \textbf{0.6888} & \textbf{0.6517} \\
@10 & EasyRec       & 0.0883 & 0.8833 & 0.1606 & 0.6853 & 0.6225 \\
    & EasyRec$^{*}$ & \textbf{0.0938} & \textbf{0.9378} & \textbf{0.1705} & \textbf{0.7696} & \textbf{0.7155} \\
    & BGE-base     & 0.0948 & 0.9478 & 0.1723 & 0.8097 & 0.7652 \\
    & BGE-base$^{*}$ & \textbf{0.0988} & \textbf{0.9878} & \textbf{0.1796} & \textbf{0.8865} & \textbf{0.8532} \\
    & KaLM          & 0.0820 & 0.8200 & 0.1491 & 0.6020 & 0.5321 \\
    & KaLM$^{*}$    & \textbf{0.0886} & \textbf{0.8856} & \textbf{0.1610} & \textbf{0.7162} & \textbf{0.6629} \\
\bottomrule
\end{tabular}
\end{table}

\begin{table}[t]
\caption{\OC{} P@1 by prompt level L1--L5 (test split, 900 queries).
Bold marks the better of each (base, adapted) pair per column.}
\label{tab:clawhub-difficulty}
\centering
\small
\setlength{\tabcolsep}{4pt}
\begin{tabular}{lccccc}
\toprule
Method & L1 & L2 & L3 & L4 & L5 \\
\midrule
EasyRec          & 0.5714 & 0.3918 & 0.4153 & 0.5351 & 0.5753 \\
EasyRec$^{*}$    & \textbf{0.6743} & \textbf{0.5614} & \textbf{0.5355} & \textbf{0.5730} & \textbf{0.6344} \\
\midrule
BGE-base        & 0.7600 & 0.6082 & 0.6120 & 0.6162 & 0.7366 \\
BGE-base$^{*}$  & \textbf{0.8686} & \textbf{0.7135} & \textbf{0.6940} & \textbf{0.7622} & \textbf{0.8118} \\
\midrule
KaLM             & 0.5714 & 0.4152 & 0.4044 & 0.4162 & 0.4624 \\
KaLM$^{*}$       & \textbf{0.7257} & \textbf{0.5146} & \textbf{0.4918} & \textbf{0.5297} & \textbf{0.5699} \\
\bottomrule
\end{tabular}
\end{table}

\begin{table}[t]
\caption{Agent-representation ablation for the unadapted EasyRec retriever on \OC{} (test split, 900 queries).
\emph{Name+Desc.}\ uses the skill name and description; \emph{All
fields} additionally appends every remaining \texttt{SKILL.md} section.}
\label{tab:ablation-repr}
\centering
\small
\setlength{\tabcolsep}{4pt}
\begin{tabular}{lcccc}
\toprule
Agent Repr. & Tok & P@1 & nDCG@10 & MRR@10 \\
\midrule
Name+Desc.  & \textbf{112} & \textbf{0.5089} & 0.6657 & \textbf{0.6116} \\
All fields  & 296 & 0.4956 & \textbf{0.6691} & 0.6106 \\
\bottomrule
\end{tabular}
\end{table}

\section{Discussion}
\label{sec:discussion}
Capability-profile supervision transfers beyond its training benchmark across
all three retriever families, though the size of the gain differs by model,
and the \OC{} results show that the signal is not limited to native-agent
catalogs. The evidence supports a shared metadata-level retrieval view, but it
does not show that routing, tool selection, and skill discovery are the same
task in a stronger operational sense.

\noindent \textbf{Model families have different query-style sensitivities.}
Table~\ref{tab:clawhub-difficulty} shows that the three retrievers respond
differently to the prompt levels. BGE-base peaks on the explicit L1 level and
remains relatively strong across all levels. All three models show a
mid-level dip at L2--L3 followed by partial recovery at L4--L5, but the
depth of the dip varies: BGE-base degrades least in the middle levels,
while EasyRec and KaLM drop more sharply at L2--L3 before recovering. These
differences persist after adaptation, suggesting that L1--L5 reflect
variation in query style rather than a single monotonic difficulty scale.

\noindent \textbf{Limitations and future work.} The present study focuses on
metadata-level retrieval and does not evaluate end-to-end task completion.
Deployment factors such as cost, latency, and safety are outside the current
scope. MuleRun agents and \OC{} skills are evaluated in separate candidate
pools; whether a single retriever can rank a mixed agent-and-skill catalog is
left to future work. Checkpoints are selected per target on a held-out split
of that target, which does not leak labels but does expose model selection to
the target distribution; a stricter protocol would select on \AS{} validation
alone. Scaling \OC{} to a larger skill catalog with human-written
queries and validating whether retrieval quality translates to downstream
task performance are the most immediate next steps.

\section{Conclusion}
Once each marketplace unit is written as a capability profile, a single
metadata-level retriever can search across agents, tools, and skills. Across
three retrieval backbones, adaptation on \AS{} improves query-to-agent
recommendation on the unseen MuleRun and \OC{} catalogs. This evidence supports
further study of unified retrieval over mixed catalogs of agents, tools, and
skills.

\bibliographystyle{ACM-Reference-Format}
\bibliography{example_paper}

\end{document}